\documentclass[a4paper]{jpconf}
\usepackage{graphicx}
\usepackage{amssymb}
\usepackage{hyperref,slashed,color,graphicx,url}
\hypersetup{colorlinks,citecolor= blue,linkcolor= blue, urlcolor=blue}

\begin{document}
\title{Neutrino and Axion Astronomy with Dark Matter Experiments}

\author{Volodymyr Takhistov}

\address{Kavli Institute for the Physics and Mathematics of the Universe (WPI), UTIAS \\The University of Tokyo, Kashiwa, Chiba 277-8583, Japan}

\ead{volodymyr.takhistov@ipmu.jp}

\begin{abstract}
Sensitive dark matter (DM) experiments can be well exploited beyond their designated targets, allowing to explore a breadth of physics topics. As we discuss, future large direct DM detection experiments constitute impressive telescopes, complementary to conventional neutrino detectors. This opens a new window into neutrino astronomy, including puzzles such as the origin of supermassive black holes and topics like supernova forecast. Furthermore, DM experiments can act as effective instruments for multimessenger astronomy. This is well illustrated by exploration of relativistic axions from transient astrophysical sources (e.g. axion star explosions), providing novel signatures as well as possible insights into the axion potential.
\end{abstract}

\section{Introduction}

The mysterious predominant constituent of matter in the Universe, dark matter (DM), is only known through gravitational interactions (see e.g. \cite{Bertone:2004pz} for review). Despite long history of exploration of possible non-gravitational interactions, DM remains elusive. Particular focus has been devoted to Weakly Interacting Massive Particles (WIMPs), whose typical $\sim$GeV - 100 TeV mass often is associated with models that can address the hierarchy problem. Many alternative particle DM candidates exist, spanning orders of magnitude in mass.

Direct DM detection aims to probe energy deposits from Galactic halo DM traversing and interacting within the experiment. With keV-level thresholds, nuclear scattering is effective for testing interactions of DM with mass above GeV~(e.g.~\cite{Cushman:2013zza,Gelmini:2018ogy}). Lighter DM can instead be efficiently studied through electron interactions~(e.g.~\cite{Essig:2011nj,Graham:2012su,Trickle:2019nya}). See also e.g.~\cite{Gelmini:2020kcu,Lawson:2019brd,Gelmini:2020xir} searches.

Data from direct detection can be analyzed through complementary halo-dependent or halo-independent approaches. In the former, assuming local DM distribution, regions of DM parameter space can be mapped in terms of mass and cross-section for a particular DM interaction. Halo-independent analysis~(e.g.~\cite{Fox:2010bz,DelNobile:2013cva}) allows to avoid uncertainties associated with the knowledge of local DM halo and instead infer local DM distribution. Recently, halo-independent method has been also applied to electron-scattering~\cite{Chen:2021qao}.

Planned large DM experiments, especially those based on argon~\cite{DarkSide-20k:2017zyg} and xenon~\cite{XENON:2015gkh,LZ:2015kxe}, can achieve multi-ton-scale fiducial volumes and have the potential to probe WIMPs with unprecedented sensitivity. Future experiments like Argo~\cite{DarkSide-20k:2017zyg} and Darwin~\cite{DARWIN:2016hyl} are poised to reach hundreds of ton-years in exposure. With increased sensitivity for exploring DM, large-scale DM experiments will eventually encounter irreducible neutrino background. Sensitivity to neutrino sources enables such DM experiments to act as effective neutrino telescopes.

\section{Neutrino Astronomy with Dark Matter Experiments}

\begin{figure}[t]
\begin{minipage}{0.42\textwidth}
\includegraphics[width=1\textwidth]{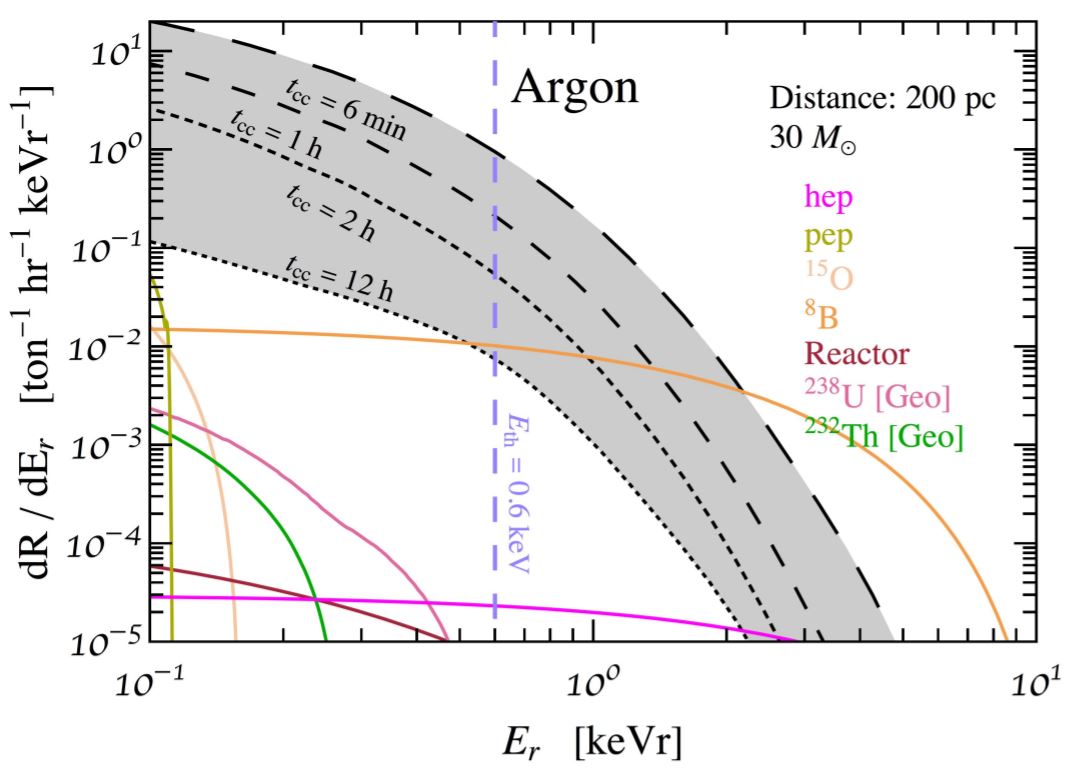}
\end{minipage}\hspace{10mm}%
\begin{minipage}{0.48\textwidth}\vspace{-2mm}
\includegraphics[width=1\textwidth]{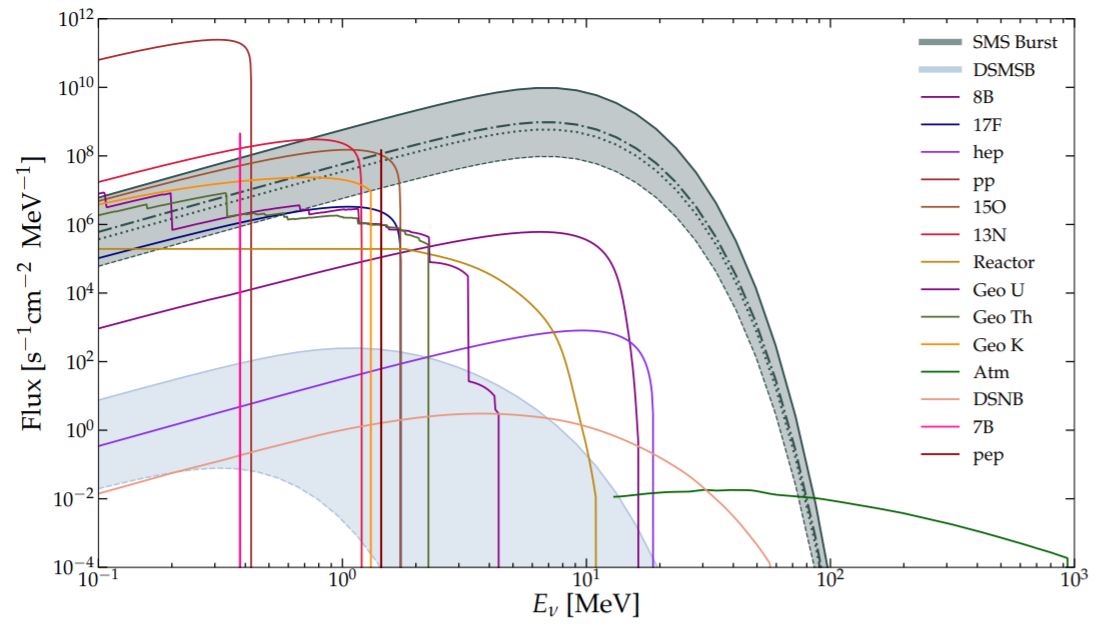}
\end{minipage} \vspace{-2mm}
\caption{ \label{fig:snsmsnu}
[Left]   Nuclear recoil spectra for pre-SN
neutrino signal in Argo-type experiment for various time intervals until
core-collapse, for
$30 M_{\odot}$ at 200 pc distance (Betelgeuse). From~\cite{Raj:2019wpy}.
 [Right] Neutrino flux from a single SMS burst and diffuse SMS background (DSMSB). The SMS burst band details signals from $\sim 10^5 M_{\odot}$ collapsing at 100 kpc (solid), 
 1 Mpc (dashed) and with rotation or magnetic field enhancements
(dash-dotted). The DSMSB band depicts variation in redshift distribution models. Background neutrino sources are shown. From~\cite{Munoz:2021sad}.
} 
\end{figure}

Variety of sources contribute to the irreducible ``neutrino floor'' background of large DM experiments~(e.g.~\cite{Billard:2013qya,Gelmini:2018ogy}), including solar (pp, pep, hep, $^7$Be, $^8$B, $^{13}$N, $^{15}$O, $^{17}$F), reactor (U, Pu), geo- (K, Th, U), atmospheric as well as diffuse supernovae background neutrinos. Degeneracy between neutrinos and DM signatures can hinder sensitivity reach.

Neutrino interactions occur through coherent elastic neutrino-nucleus scattering (CE$\nu$NS), where neutrinos with energies $\lesssim 50$ MeV interact with nucleus as a whole. Recently observed~\cite{COHERENT:2017ipa}, CE$\nu$NS are sensitive to all six $(\nu_e, \overline{\nu}_e, \nu_{\mu}, \overline{\nu}_{\mu}, \nu_{\tau}, \overline{\nu}_{\tau})$ neutrino flavors and are unconstrained by kinematic thresholds. Large DM experiments with heavy targets can exploit CE$\nu$NS as respective cross-section scales with the number of nucleons $\sim N^2$. Hence, large DM experiments can serve as effective telescopes complementary to conventional neutrino detectors. Boasting low keV-level thresholds, DM experiments can explore a breadth of topics (e.g. geoneutrinos~\cite{Gelmini:2018gqa}, new neutrino interactions~\cite{Harnik:2012ni}) and open a new window into neutrino astronomy, as we illustrate.

Stars with mass $\gtrsim 8 M_{\odot}$ at the end of thermonuclear evolution lifetime undergo a violent core-collapse supernova (SN) explosion. Copious accompanying emission of $\sim 10 - 30$ MeV neutrinos within a few second burst can carry away
$\sim 10^{53}$ erg of the star’s gravitational binding
energy, confirmed by historic observations of SN 1987A. Hunt for next Galactic SN is a major target~\cite{Mirizzi:2015eza} and unique perspective can be offered by large DM experiments~(e.g.~\cite{Lang:2016zhv}). Prior to core-collapse, there is significant emission of $\sim$MeV ``pre-supernova'' neutrinos associated with final nuclear burning stages~(e.g.~\cite{Kato:2015faa,Patton:2017neq,Odrzywolek:2003vn}). Detecting such neutrinos can offer unique opportunity to ``forecast'' SN and gain insights into the final stages of star's lifetime. Intriguingly, large DM experiments can readily observe pre-supernova neutrinos~\cite{Raj:2019wpy}. Considering the well studied near-by red supergiant Betelgeuse for reference, on Fig.~\ref{fig:snsmsnu} we depict the resulting nuclear-recoil spectra expected in Argo-type experiment from a pre-SN neutrino signal. This offers complementary perspectives to conventional detectors~\cite{Super-Kamiokande:2019xnm}, allowing all-flavor sensitivity.

DM experiments can also offer new perspectives on central puzzles such as the origin of supermassive $\sim 10^6-10^9 M_{\odot}$ black holes (SMBHs), residing within galactic centers~\cite{vanderMarel:1997hr} and powering quasars and Active Galactic Nuclei (AGN)~\cite{rees:1984}. Variety of standard astrophysical pathways for formation~\cite{begelman:11978} go through a collapsing supermassive $\gtrsim 3 \times 10^4 M_{\odot}$ star (SMS) phase, resulting in seed black holes for SBMHs. Emitted prodigious neutrino flux accompanying SMS collapse~\cite{Shi:1998jx,Shi:1998nd,Linke:2001mq}, orders above that of SN, can provide essential insights into SMBH formation mechanisms. Low-threshold large DM experiments can effectively capture these low-energy neutrinos. While the rates and redshift SMS distribution is unknown, collapsing SMS neutrino bursts can be well seen up to Andromeda $(\sim 1)$Mpc distances by future DM experiments, as displayed on Fig.~\ref{fig:snsmsnu}. Diffuse neutrino background from historic SMS collapses (see Fig.~\ref{fig:snsmsnu}) could also be in principle observable.

\begin{figure*}[t]
\centering
\includegraphics[width=0.8\textwidth]{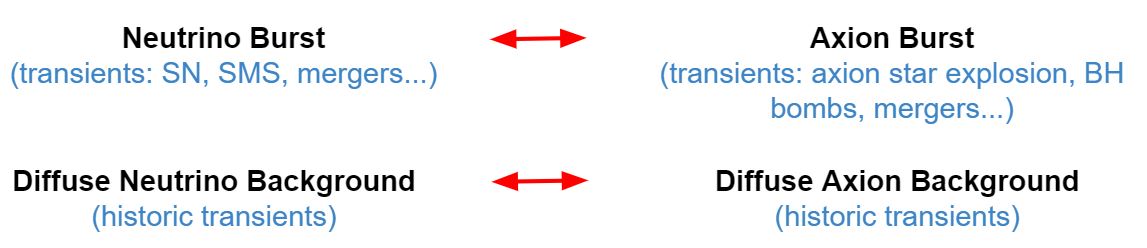}
\vspace{-2mm}
\caption{\label{fig:axionastro} Schematic of axion astronomy, a neutrino analogy.}
\end{figure*}

\section{Axion Astronomy with Dark Matter Experiments}

Axions constitute another attractive class of DM candidates. The QCD axion provides a leading solution to the strong CP problem~\cite{peccei:1977,weinberg:1978,wilczek:1978} and axions/axion-like particles are ubiquitous in theories of extra dimensions and string theories~(e.g.~\cite{witten:1984,Arvanitaki:2009fg}).  
While experimental searches have been primarily devoted to non-relativistic axions contributing to cold DM~(e.g.~\cite{Irastorza:2018dyq}), relativistic axions can appear from a variety of sources. This includes early Universe production mechanisms contributing to ``cosmic axion background''~\cite{Dror:2021nyr}, as well as continuous production from astrophysical sources like stars~\cite{Raffelt:2006cw}.

Interesting relativistic axion signatures can be associated with transient astrophysical sources~\cite{Eby:2021ece}. More generally, we can outline ``axion astronomy'' associated with transient phenomena exploiting direct analogy with neutrino astronomy (see e.g. Section~2), as depicted in Fig.~\ref{fig:axionastro}. Similar to neutrino bursts from transients like SN, SMS and neutron star mergers, bursts of relativistic axions can be associated with transients like axion star (bosenova) explosions~\cite{Eby:2016cnq,Helfer:2016ljl,Levkov:2016rkk}, black hole supperradiance ``bombs''~(e.g.~\cite{Baumann:2018vus,Arvanitaki:2014wva}) and SN~\cite{Raffelt:1999tx}. Historic contributions of transients will establish ``diffuse axion background''.

\begin{figure}[t] 
\begin{minipage}{0.49\textwidth}
\includegraphics[width=1\textwidth]{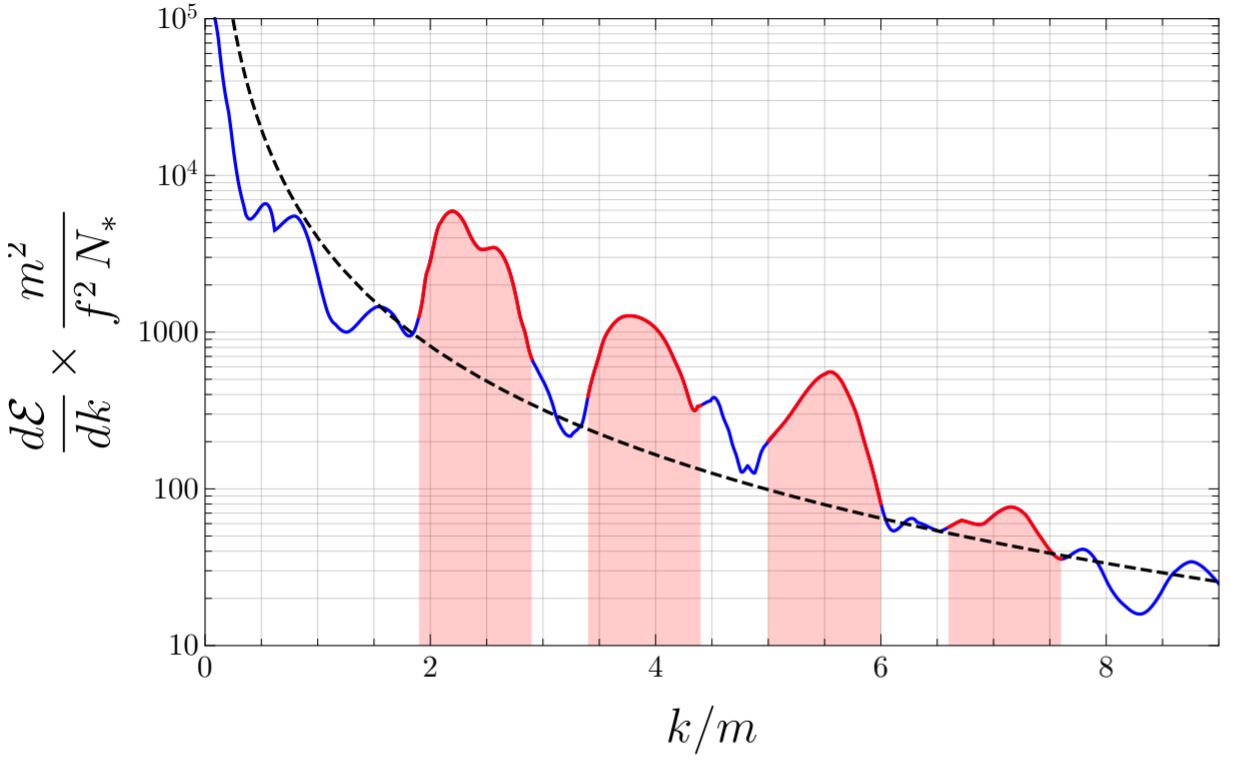}
\end{minipage}\hspace{10mm}%
\begin{minipage}{0.49\textwidth}
\includegraphics[width=0.9\textwidth]{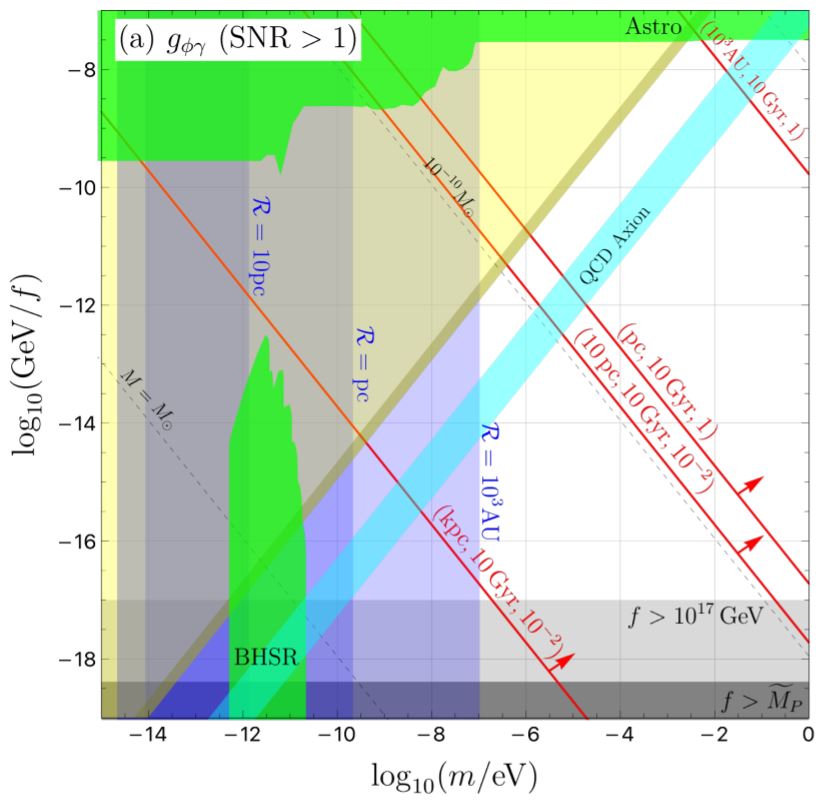}
\end{minipage} \vspace{-2mm}
\caption{\label{fig:axionstar} [Left] Spectrum of emitted axions from axion star explosions, from~\cite{Levkov:2016rkk}. Peaks are shaded in red and N$_*$ denotes the number of explosions during the collapse. [Right] Sensitivity reach of ABRACADABRA to axion star bosenova (blue regions), with existing constraints and a choice of (distance, explosion timescale, DM fraction in axion stars). Yellow regions indicate where parametric
resonance conversion of axions to photons would be relevant.   From~\cite{Eby:2021ece}.
} 
\end{figure}

Detection prospects for transient axions can be readily demonstrated with explosions of axion stars~\cite{Eby:2021ece}. Such condensed bound gravitational configurations can appear from the early Universe e.g. within miniclusters~\cite{Kolb:1993zz,Eggemeier:2019jsu}.  Fig.~\ref{fig:axionstar} depicts reproduced relativistic axion emission spectrum in terms of momentum over mass $k/m$ from detailed numerical simulations~\cite{Levkov:2016rkk}, assuming QCD axion potential with self-interactions. The characteristic~\cite{Eby:2014fya} critical mass of stable axion star and minimum radius are $M_c \sim 10^{-11}M_{\odot} f_{12}/m_5$ and $R_c \sim 100~\textrm{km}/(f_{12}m_5)$, respectively, in terms of mass $m_5 = m/(10^{-5}\textrm{eV})$ and axion decay constant $f_{12} = f/(6 \times 10^{11} \textrm{GeV})$. 

In contrast to ambient axion DM searches, an explosive axion burst of duration $\delta t \sim 30$~ns/$m_5$ emitting significant fraction of initial axion star mass with an integrated main peak emission of $\mathcal{E} \sim 10^{41}~\textrm{GeV}f_{12}^2/m_5$~\cite{Levkov:2016rkk}, assuming minimum wave spreading, can lead to significant axion energy densities in the detector. In case of derivative axion coupling, velocity enhancement $v_{\ast}/v_{\rm DM} \sim 10^3$ compared to non-relativistic DM is also possible.
Taking into account effects of coherence and wave spatio-temporal spreading, Fig.~\ref{fig:axionstar} depicts sensitivity to axion star bursts for ABRACADABRA-type experiment~\cite{Kahn:2016aff} for axion-photon coupling. While sensitivity of conventional axion DM searches falls off as $1/f$ or faster due to axion coupling, signals from axion star explosions are enhanced as $\sim f$ from $\sqrt{\mathcal{E}} \sim f$. Thus, axion star bursts may offer a preferred method of discovery for large values of the decay constant $f$. Since the axion burst spectrum is sensitive to self-interactions, relativistic axions from transient sources pave a way for spectroscopy of the fundamental axion potential, challenging otherwise.
Aside new contributions to multimessenger signatures~\cite{Tkachev:2014dpa}, diffuse axion background from historic transients establishes yet another class of observables.
 
\section{Summary}

New ideas for DM experiments dramatically expand the scope of physics being explored. Future large DM experiments constitute effective neutrino telescopes, complementary to conventional neutrino experiments, opening a new window into neutrino astronomy. Drawing on lessons from neutrinos, axion astronomy with transient sources opens a completely novel direction to explore axions, including the fundamental axion potential with experiments as well as multimessenger signatures.

\section*{Acknowledgements}
The author would like to thank the organizers of the 17th International Conference on Topics in Astroparticle and Underground Physics (TAUP-2021) for
the opportunity to present these results, as well as collaborators of discussed studies and Graciela B. Gelmini for comments. This work was supported in part by the World Premier International Research Center Initiative (WPI), MEXT, Japan.

\section*{References}

\bibliography{procbib.bib}

\providecommand{\newblock}{}
\begin{thebibliography}{10}
\expandafter\ifx\csname url\endcsname\relax
  \def\url#1{{\tt #1}}\fi
\expandafter\ifx\csname urlprefix\endcsname\relax\def\urlprefix{URL }\fi
\providecommand{\eprint}[2][]{\url{#2}}

\bibitem{Bertone:2004pz}
Bertone G, Hooper D and Silk J 2005 {\em Phys. Rept.\/} {\bf 405} 279--390
  (\textit{Preprint}  \href{https://arxiv.org/abs/hep-ph/0404175}{{\ttfamily
  hep-ph/0404175}})

\bibitem{Cushman:2013zza}
Cushman P {\em et~al.\/} 2013 {\em {Community Summer Study 2013}: {Snowmass on
  the Mississippi}\/} (\textit{Preprint}
  \href{https://arxiv.org/abs/1310.8327}{{\ttfamily 1310.8327}})

\bibitem{Gelmini:2018ogy}
Gelmini G~B, Takhistov V and Witte S~J 2018 {\em JCAP\/} {\bf 07} 009 [Erratum:
  JCAP 02, E02 (2019)] (\textit{Preprint}
  \href{https://arxiv.org/abs/1804.01638}{{\ttfamily 1804.01638}})

\bibitem{Essig:2011nj}
Essig R, Mardon J and Volansky T 2012 {\em Phys. Rev. D\/} {\bf 85} 076007
  (\textit{Preprint}  \href{https://arxiv.org/abs/1108.5383}{{\ttfamily
  1108.5383}})

\bibitem{Graham:2012su}
Graham P~W, Kaplan D~E, Rajendran S and Walters M~T 2012 {\em Phys. Dark
  Univ.\/} {\bf 1} 32--49 (\textit{Preprint}
  \href{https://arxiv.org/abs/1203.2531}{{\ttfamily 1203.2531}})

\bibitem{Trickle:2019nya}
Trickle T, Zhang Z, Zurek K~M, Inzani K and Griffin S 2020 {\em JHEP\/} {\bf
  03} 036 (\textit{Preprint}
  \href{https://arxiv.org/abs/1910.08092}{{\ttfamily 1910.08092}})

\bibitem{Gelmini:2020kcu}
Gelmini G~B, Millar A~J, Takhistov V and Vitagliano E 2020 {\em Phys. Rev. D\/}
  {\bf 102} 043003 (\textit{Preprint}
  \href{https://arxiv.org/abs/2006.06836}{{\ttfamily 2006.06836}})

\bibitem{Lawson:2019brd}
Lawson M, Millar A~J, Pancaldi M, Vitagliano E and Wilczek F 2019 {\em Phys.
  Rev. Lett.\/} {\bf 123} 141802 (\textit{Preprint}
  \href{https://arxiv.org/abs/1904.11872}{{\ttfamily 1904.11872}})

\bibitem{Gelmini:2020xir}
Gelmini G~B, Takhistov V and Vitagliano E 2020 {\em Phys. Lett. B\/} {\bf 809}
  135779 (\textit{Preprint}  \href{https://arxiv.org/abs/2006.13909}{{\ttfamily
  2006.13909}})

\bibitem{Fox:2010bz}
Fox P~J, Liu J and Weiner N 2011 {\em Phys. Rev. D\/} {\bf 83} 103514
  (\textit{Preprint}  \href{https://arxiv.org/abs/1011.1915}{{\ttfamily
  1011.1915}})

\bibitem{DelNobile:2013cva}
Del~Nobile E, Gelmini G, Gondolo P and Huh J~H 2013 {\em JCAP\/} {\bf 10} 048
  (\textit{Preprint}  \href{https://arxiv.org/abs/1306.5273}{{\ttfamily
  1306.5273}})

\bibitem{Chen:2021qao}
Chen M, Gelmini G~B and Takhistov V 2021   (\textit{Preprint}
  \href{https://arxiv.org/abs/2105.08101}{{\ttfamily 2105.08101}})

\bibitem{DarkSide-20k:2017zyg}
Aalseth C~E {\em et~al.\/} (DarkSide-20k) 2018 {\em Eur. Phys. J. Plus\/} {\bf
  133} 131 (\textit{Preprint}
  \href{https://arxiv.org/abs/1707.08145}{{\ttfamily 1707.08145}})

\bibitem{XENON:2015gkh}
Aprile E {\em et~al.\/} (XENON) 2016 {\em JCAP\/} {\bf 04} 027
  (\textit{Preprint}  \href{https://arxiv.org/abs/1512.07501}{{\ttfamily
  1512.07501}})

\bibitem{LZ:2015kxe}
Akerib D~S {\em et~al.\/} (LZ) 2015   (\textit{Preprint}
  \href{https://arxiv.org/abs/1509.02910}{{\ttfamily 1509.02910}})

\bibitem{DARWIN:2016hyl}
Aalbers J {\em et~al.\/} (DARWIN) 2016 {\em JCAP\/} {\bf 11} 017
  (\textit{Preprint}  \href{https://arxiv.org/abs/1606.07001}{{\ttfamily
  1606.07001}})

\bibitem{Raj:2019wpy}
Raj N, Takhistov V and Witte S~J 2020 {\em Phys. Rev. D\/} {\bf 101} 043008
  (\textit{Preprint}  \href{https://arxiv.org/abs/1905.09283}{{\ttfamily
  1905.09283}})

\bibitem{Munoz:2021sad}
Munoz V, Takhistov V, Witte S~J and Fuller G~M 2021   (\textit{Preprint}
  \href{https://arxiv.org/abs/2102.00885}{{\ttfamily 2102.00885}})

\bibitem{Billard:2013qya}
Billard J, Strigari L and Figueroa-Feliciano E 2014 {\em Phys. Rev. D\/} {\bf
  89} 023524 (\textit{Preprint}
  \href{https://arxiv.org/abs/1307.5458}{{\ttfamily 1307.5458}})

\bibitem{COHERENT:2017ipa}
Akimov D {\em et~al.\/} (COHERENT) 2017 {\em Science\/} {\bf 357} 1123--1126
  (\textit{Preprint}  \href{https://arxiv.org/abs/1708.01294}{{\ttfamily
  1708.01294}})

\bibitem{Gelmini:2018gqa}
Gelmini G~B, Takhistov V and Witte S~J 2019 {\em Phys. Rev. D\/} {\bf 99}
  093009 (\textit{Preprint}  \href{https://arxiv.org/abs/1812.05550}{{\ttfamily
  1812.05550}})

\bibitem{Harnik:2012ni}
Harnik R, Kopp J and Machado P~A~N 2012 {\em JCAP\/} {\bf 07} 026
  (\textit{Preprint}  \href{https://arxiv.org/abs/1202.6073}{{\ttfamily
  1202.6073}})

\bibitem{Mirizzi:2015eza}
Mirizzi A, Tamborra I, Janka H~T, Saviano N, Scholberg K, Bollig R, Hudepohl L
  and Chakraborty S 2016 {\em Riv. Nuovo Cim.\/} {\bf 39} 1--112
  (\textit{Preprint}  \href{https://arxiv.org/abs/1508.00785}{{\ttfamily
  1508.00785}})

\bibitem{Lang:2016zhv}
Lang R~F, McCabe C, Reichard S, Selvi M and Tamborra I 2016 {\em Phys. Rev.
  D\/} {\bf 94} 103009 (\textit{Preprint}
  \href{https://arxiv.org/abs/1606.09243}{{\ttfamily 1606.09243}})

\bibitem{Kato:2015faa}
Kato C, Azari M~D, Yamada S, Takahashi K, Umeda H, Yoshida T and Ishidoshiro K
  2015 {\em Astrophys. J.\/} {\bf 808} 168 (\textit{Preprint}
  \href{https://arxiv.org/abs/1506.02358}{{\ttfamily 1506.02358}})

\bibitem{Patton:2017neq}
Patton K~M, Lunardini C, Farmer R~J and Timmes F~X 2017 {\em Astrophys. J.\/}
  {\bf 851} 6 (\textit{Preprint}
  \href{https://arxiv.org/abs/1709.01877}{{\ttfamily 1709.01877}})

\bibitem{Odrzywolek:2003vn}
Odrzywolek A, Misiaszek M and Kutschera M 2004 {\em Astropart. Phys.\/} {\bf
  21} 303--313 (\textit{Preprint}
  \href{https://arxiv.org/abs/astro-ph/0311012}{{\ttfamily astro-ph/0311012}})

\bibitem{Super-Kamiokande:2019xnm}
Simpson C {\em et~al.\/} (Super-Kamiokande) 2019 {\em Astrophys. J.\/} {\bf
  885} 133 (\textit{Preprint}
  \href{https://arxiv.org/abs/1908.07551}{{\ttfamily 1908.07551}})

\bibitem{vanderMarel:1997hr}
van~der Marel R~P, de~Zeeuw P~T, Rix H~W and Quinlan G~D 1997 {\em Nature\/}
  {\bf 385} 610 (\textit{Preprint}
  \href{https://arxiv.org/abs/astro-ph/9702106}{{\ttfamily astro-ph/9702106}})

\bibitem{rees:1984}
Rees M~J 1984 {\em Annual Review of Astronomy and Astrophysics\/} {\bf 22}
  471--506 \urlprefix\url{https://doi.org/10.1146/annurev.aa.22.090184.002351}

\bibitem{begelman:11978}
Begelman M~C and Rees M~J 1978 {\em Monthly Notices of the Royal Astronomical
  Society\/} {\bf 185} 847--860 ISSN 0035-8711
  \urlprefix\url{https://doi.org/10.1093/mnras/185.4.847}

\bibitem{Shi:1998jx}
Shi X~D, Fuller G~M and Halzen F 1998 {\em Phys. Rev. Lett.\/} {\bf 81}
  5722--5725 (\textit{Preprint}
  \href{https://arxiv.org/abs/astro-ph/9805242}{{\ttfamily astro-ph/9805242}})

\bibitem{Shi:1998nd}
Shi X~D and Fuller G~M 1998 {\em Astrophys. J.\/} {\bf 503} 307
  (\textit{Preprint}  \href{https://arxiv.org/abs/astro-ph/9801106}{{\ttfamily
  astro-ph/9801106}})

\bibitem{Linke:2001mq}
Linke F, Font J~A, Janka H~T, Muller E and Papadopoulos P 2001 {\em Astron.
  Astrophys.\/} {\bf 376} 568 (\textit{Preprint}
  \href{https://arxiv.org/abs/astro-ph/0103144}{{\ttfamily astro-ph/0103144}})

\bibitem{peccei:1977}
Peccei R~D and Quinn H~R 1977 {\em Phys. Rev. Lett.\/} {\bf 38}(25) 1440--1443
  \urlprefix\url{https://link.aps.org/doi/10.1103/PhysRevLett.38.1440}

\bibitem{weinberg:1978}
Weinberg S 1978 {\em Phys. Rev. Lett.\/} {\bf 40}(4) 223--226
  \urlprefix\url{https://link.aps.org/doi/10.1103/PhysRevLett.40.223}

\bibitem{wilczek:1978}
Wilczek F 1978 {\em Phys. Rev. Lett.\/} {\bf 40}(5) 279--282
  \urlprefix\url{https://link.aps.org/doi/10.1103/PhysRevLett.40.279}

\bibitem{witten:1984}
Witten E 1984 {\em Physics Letters B\/} {\bf 149} 351--356 ISSN 0370-2693
  \urlprefix\url{https://www.sciencedirect.com/science/article/pii/0370269384904222}

\bibitem{Arvanitaki:2009fg}
Arvanitaki A, Dimopoulos S, Dubovsky S, Kaloper N and March-Russell J 2010 {\em
  Phys. Rev.\/} {\bf D81} 123530 (\textit{Preprint}
  \href{https://arxiv.org/abs/0905.4720}{{\ttfamily 0905.4720}})

\bibitem{Irastorza:2018dyq}
Irastorza I~G and Redondo J 2018 {\em Prog. Part. Nucl. Phys.\/} {\bf 102}
  89--159 (\textit{Preprint}
  \href{https://arxiv.org/abs/1801.08127}{{\ttfamily 1801.08127}})

\bibitem{Dror:2021nyr}
Dror J~A, Murayama H and Rodd N~L 2021 {\em Phys. Rev. D\/} {\bf 103} 115004
  (\textit{Preprint}  \href{https://arxiv.org/abs/2101.09287}{{\ttfamily
  2101.09287}})

\bibitem{Raffelt:2006cw}
Raffelt G~G 2008 {\em Lect. Notes Phys.\/} {\bf 741} 51--71 (\textit{Preprint}
  \href{https://arxiv.org/abs/hep-ph/0611350}{{\ttfamily hep-ph/0611350}})

\bibitem{Eby:2021ece}
Eby J, Shirai S, Stadnik Y~V and Takhistov V 2021   (\textit{Preprint}
  \href{https://arxiv.org/abs/2106.14893}{{\ttfamily 2106.14893}})

\bibitem{Eby:2016cnq}
Eby J, Leembruggen M, Suranyi P and Wijewardhana L~C~R 2016 {\em JHEP\/} {\bf
  12} 066 (\textit{Preprint}
  \href{https://arxiv.org/abs/1608.06911}{{\ttfamily 1608.06911}})

\bibitem{Helfer:2016ljl}
Helfer T, Marsh D~J~E, Clough K, Fairbairn M, Lim E~A and Becerril R 2017 {\em
  JCAP\/} {\bf 1703} 055 (\textit{Preprint}
  \href{https://arxiv.org/abs/1609.04724}{{\ttfamily 1609.04724}})

\bibitem{Levkov:2016rkk}
Levkov D~G, Panin A~G and Tkachev I~I 2017 {\em Phys. Rev. Lett.\/} {\bf 118}
  011301 (\textit{Preprint}  \href{https://arxiv.org/abs/1609.03611}{{\ttfamily
  1609.03611}})

\bibitem{Baumann:2018vus}
Baumann D, Chia H~S and Porto R~A 2019 {\em Phys. Rev. D\/} {\bf 99} 044001
  (\textit{Preprint}  \href{https://arxiv.org/abs/1804.03208}{{\ttfamily
  1804.03208}})

\bibitem{Arvanitaki:2014wva}
Arvanitaki A, Baryakhtar M and Huang X 2015 {\em Phys. Rev. D\/} {\bf 91}
  084011 (\textit{Preprint}  \href{https://arxiv.org/abs/1411.2263}{{\ttfamily
  1411.2263}})

\bibitem{Raffelt:1999tx}
Raffelt G~G 1999 {\em Ann. Rev. Nucl. Part. Sci.\/} {\bf 49} 163--216
  (\textit{Preprint}  \href{https://arxiv.org/abs/hep-ph/9903472}{{\ttfamily
  hep-ph/9903472}})

\bibitem{Kolb:1993zz}
Kolb E~W and Tkachev I~I 1993 {\em Phys. Rev. Lett.\/} {\bf 71} 3051--3054
  (\textit{Preprint}  \href{https://arxiv.org/abs/hep-ph/9303313}{{\ttfamily
  hep-ph/9303313}})

\bibitem{Eggemeier:2019jsu}
Eggemeier B and Niemeyer J~C 2019 {\em Phys. Rev. D\/} {\bf 100} 063528
  (\textit{Preprint}  \href{https://arxiv.org/abs/1906.01348}{{\ttfamily
  1906.01348}})

\bibitem{Eby:2014fya}
Eby J, Suranyi P, Vaz C and Wijewardhana L~C~R 2015 {\em JHEP\/} {\bf 03} 080
  [Erratum: JHEP 11, 134 (2016)] (\textit{Preprint}
  \href{https://arxiv.org/abs/1412.3430}{{\ttfamily 1412.3430}})

\bibitem{Kahn:2016aff}
Kahn Y, Safdi B~R and Thaler J 2016 {\em Phys. Rev. Lett.\/} {\bf 117} 141801
  (\textit{Preprint}  \href{https://arxiv.org/abs/1602.01086}{{\ttfamily
  1602.01086}})

\bibitem{Tkachev:2014dpa}
Tkachev I~I 2015 {\em JETP Lett.\/} {\bf 101} 1--6 (\textit{Preprint}
  \href{https://arxiv.org/abs/1411.3900}{{\ttfamily 1411.3900}})

\end{thebibliography}
\bibliographystyle{iopart-num.bst}

\end{document}